# Broad 1$^{st}$ Order Phase Transitions: phase coexistence, unusual hysteresis and glasslike 'kinetic arrest'.

P Chaddah*
RR Centre for Advanced Technology, Indore 452013, India

*We discuss the phenomenology of phase transitions studied with two control variables. Such measurements have become routine with magnetic field and temperature being varied for 1$^{st}$ order magnetic transitions with an ease not conceivable with pressure and temperature. Similar ease may be possible with electric field as the second control variable, as in current studies on multiferroics. We develop this phenomenology for broad 1$^{st}$ order transitions that occur over a range of the control variable, as is common in substitutional alloys with inherent quenched disorder, including the possibility of a glasslike arrest of the transition occurring within this range of either control variable. The interrupted transition then results in the two competing phases coexisting to the lowest temperature. The new experimental protocol with acronym CHUF, where the sample is cooled and heated under different values of the second control variable, allows a detailed study of glasslike arrest of kinetics of the phase transformation. The many experimental studies on 1$^{st}$ order magnetic transitions in a large number of materials, that provide the raison d'etre for this phenomenology, shall be discussed subsequently.*

## Introduction

The Ehrenfest classification of phase transitions covers the possibility of a phase transition being of any integer order. The classification is mathematically rigorous, in that some integer derivative of the free energy has to be discontinuous while one lower derivative has to be continuous for a transition of that integer order. Some phase transitions that were, as can be expected mathematically, falling outside its classification scheme of integer orders have been discovered. This treatment of phase transitions is also not very illuminating on why (or whether) one cannot supercool or superheat across 2$^{nd}$ and higher order transitions. Observation of latent heat is an experimental requirement for classifying a phase transition as 1$^{st}$ order; compliance with the Clausius-Clapeyron relation provides conclusive evidence. A major point in this classification is that temperature and the other control variable (pressure, or magnetic field) appear symmetrically.

In what is often termed the 'modern classification' there are only two types of phase transitions; 1$^{st}$ order are defined consistent with the Ehrenfest classification while all other phase transitions (including those exceptions that fall outside the Ehrenfest classification) are classified as 2$^{nd}$ order. The modern classification is based on Landau theory in which the free energy is written in a power series expansion of an 'order parameter' keeping terms of low order, as required by physical constraints. As brought out in the schematic figures 1 & 2, this formulation explains that one can supercool or superheat only up to a limit across $1^{st}$ order transitions, because multiple local minima in free energy exist only in a finite window $T^* < T < T^{**}$. It also explains that one cannot supercool or superheat across $2^{nd}$ order transitions because there is always only one local minimum of free energy in these cases. This result is most important for our discussion; this formulation has provided legitimacy to hysteresis being used as an experimental signature of a $1^{st}$ order transition. However, the 'modern classification' predominantly treats temperature as the control variable and does not treat the second control variable symmetrically; discussions on the second control variable became very necessary because of the many experimental studies on $1^{st}$ order magnetic transitions in magnetic functional materials where the magnetic field H is routinely used to cause the phase transition. Specifically, we wish to emphasize that f(S=0,T) is justifiably treated as independent of T. If, however, f is expanded in terms of a second control variable (say V), then f(S=0,V) cannot be assumed to be independent of V. This is naively obvious for the second control variable being pressure P, since change in P will change inter-atomic distances and hence the energy.

## Varying two control variables: supercooled and arrested states

We start with the initial such effort by Chaddah and Roy [1] to consider the second control variable, where they considered a pressure dependence (or density dependence) in the free energy expansion. Their phenomenology for the calculation of the limit of supercooling T*(P) and the limit of superheating T**(P) showed that a constant-P variation of T results in a hysteresis ΔT(P) = [T**(P) – T*(P)] whose variation with P is qualitatively dictated by the variation of $T_C(P)$. If $T_C(P)$ falls as P rises then ΔT(P) rises as P rises; whereas if $T_C(P)$ rises as P rises then ΔT(P) falls as P rises (see schematic figure 3 from [2]). They conjectured that the same dependence may hold if H is the second control variable. Violations of this conjecture have not been found so far [3], indicating that a phenomenological justification should be looked for! A second question they addressed was whether causing the phase transition by varying P allows supercooling till the T*(P) line. They argued that in the presence of any quenched disorder such variation of the second control variable produces fluctuations that terminate supercooling before T* is reached; the limit of supercooling is best approached (to a temperature above T* where the barrier height, see fig 1, has reduced to near kT) when only T is varied [4].

A very different question that was addressed through these studies on the use of two control variables was the effect of the proximity of the (T*,H*) line on glass-like arrest of kinetics. [*Measurements with varying H are experimentally much easier than measurements with varying P because H is transmitted in vacuum while P requires a medium, and that interferes with control of temperature, especially at low T. In view of the many experimental studies on $1^{st}$ order magnetic transitions in a large number of materials that provide the raison d'etre for this*

*phenomenology, we shall now consider H as the second control variable. H can be replaced by P (or E) with FM/AFM replaced by high density phase/low density phase (or high dielectric constant phase/low dielectric constant phase).*] It is clear from figure 1 that the barrier separating the metastable disordered state and the stable ordered state reduces as T (or H) is varied from $T_C$ towards $T^*$. If the metastable state is ergodic on the experimental time scale, its relaxation towards equilibrium is dictated by this barrier height and relaxation rate will increase. As one approaches the glass-like kinetically arrested state (at $T_k$ or at $H_k$), non-ergodicity sets in and the system stops exploring phase space and the barrier height becomes less important than 'diffusivity' in determining relaxation rate [5]. In this case the relaxation rate is dictated by the (in-)ability to explore phase space and it rises as 'kinetic arrest line' ($T_k$,$H_k$) is approached. (This increase with lowering T follows the KWW [Kohlrausch-Williams-Watt] law, and these time dependent measurements are carried out at various temperatures to confirm the onset of a glass-like arrested state [6].) These contrasting behaviours were pointed out while trying to simultaneously understand glass-like kinetic arrest and supercooling in the context of the free energy expansion in Landau theory. It should be noted that standard treatments of glass-formation deal with the potential energy landscape and not the sample free energy. In that scenario also glass formation occurs when processes that explore a reasonable neighbourhood in configuration space are inhibited, and the system is trapped in a deep potential well. In this treatment the glass-temperature $T_g$ is not compared with $T^*$ (but is sometimes compared with $T_C$). The new understanding attempted was that if $T_k$ was higher than $T^*$ then non-ergodicity sets in while the system is in the metastable state, and a kinetically-arrested-glass will form even at normal cooling rates. If $T^*$ was higher than $T_k$ then the metastable-to-stable transformation would occur at normal cooling rates, and rapid cooling becomes essential for glass-like arrest of the $1^{st}$ order transition. Magnetic first order transitions are considered in figure 4; (a) represents a ferromagnetic (FM) to antiferromagnetic (AFM) transition with lowering T, while (c) represents an AFM to FM transition with lowering T. Figure 4 brings out what could happen on cooling at different values of the second control variable; the system evolves from $T_k < T^*$ to $T_k > T^*$ as the second control variable H is varied. (The sign of the slopes of ($H^*$,$T^*$) and ($H^{**}$,$T^{**}$) lines is dictated by the requirement of Le Chatelier's Principle that isothermal increase of H must cause a transition from the AFM phase to the FM phase.) As we have emphasized repeatedly, one can hope to go from a metglass-like situation to an O-terphenyl-like situation in the same material [2]! In figure 4(a) this happens with rising H as $T_k < T^*$ changes to $T_k > T^*$, while in the case of figure 4(c) this happens with reducing H. We shall now discuss some interesting thermomagnetic history effects that can be visualized, and some of which have been established experimentally.

# Thermomagnetic history effects: contrasting supercooled and arrested states

We consider a point with coordinates ($T_1$,$H_1$) as shown in the schematic of a high-T FM phase transforming to a low-T AFM phase on cooling, with $T_k$ > T* at large H. We consider two paths of reaching it from above the phase transition. In path A the transition to the AFM phase is completed, while in path C the FM phase is kinetically arrested at ($T_1$,$H_2$) and remains arrested as H is isothermally reduced to $H_1$. This is because the temperature remains below $T_k(H)$ in this isothermal reduction of H. We have a *path-dependent state* at ($T_1$,$H_1$).

After reaching ($T_1$,$H_1$) by following path C, we now raise T in constant field $H_1$. The kinetically arrested FM state *devitrifies* to the equilibrium AFM phase when $T_k(H_1)$ is crossed (see figure 4(b)). This is because T is below T*, there is no barrier, and the system is able to explore phase space since T is above $T_k$ As T is raised further, the equilibrium AFM phase transforms to the equilibrium FM phase at T**($H_1$). *This re-entrant FM to AFM to FM transition on warming in a suitable constant H* is seen in the 'Cooling and Heating in Unequal Field' (CHUF) protocol, and will not be seen in the absence of the $T_k(H)$ line; it confirms that the underlying first order transition was (glass-like) kinetically arrested during cooling in $H_2$. CHUF is a new protocol (which was introduced by me and my collaborators [7]) where the use of the second control variable can provide an experimentally much easier method (c.f. the variation of relaxation rate with decreasing T [6]) to confirm the formation of a glass-like arrested state.

A similar reentrant 'devitrification' of arrested AFM to equilibrium FM followed by 'melting' of equilibrium FM to equilibrium AFM will be seen in the schematic of figure 4(c), except that the CHUF protocol for this case requires that the cooling field $H_2$ should be *lower* than the warming field $H_1$. The reentrant transition under CHUF protocol is thus seen only for $H_2$ > $H_1$ in figure 4(b), and only for $H_2$ < $H_1$ in figure 4(c). This is an important qualitative difference for the two cases. This CHUF protocol was proposed and exploited for magnetic 1st order transitions; its extension to the CHUP protocol for 1st order structural transitions was proposed by Chaddah and Banerjee [7] in 2011, and an extension (to a CHUE protocol) for 1st order dielectric transitions is obvious. The utilization of the second thermodynamic control variable in the above (CHUF and its analogues) is conceptual, and will become more striking (even visually!) when we discuss disorder-broadened 1st order transitions.

We now consider the manifestation of the two schematics [figures 5(a) and (b)] in the oft-measured isothermal M vs. H, after cooling to a low temperature $T_0$ (in H=0) following different paths. In both cases the H=0 state is an AFM state which is the equilibrium phase in case (a), and the 'glass-like kinetically arrested' high-T phase in case (b). On increasing H to a suitably large value, there is a 1st order transition to the FM phase at H** in case (a), and a

'devitrification' to the equilibrium FM phase at $H_k$ in case (b). In both cases the subsequent reduction of H to 0 leaves the system in the FM phase because the system does not simultaneously goes below $T^*(H)$ and above $T_k(H)$ in case (a). This manifestation of kinetic arrest *continues to be seen at all lower temperatures*. Unlike the behavior under the CHUF protocol, the qualitative behavior corresponds to an initial AFM state in the virgin cycle, and a final FM state in the hysteresis cycle. The consequent observation of a virgin curve lying outside the envelope hysteresis curve (or an open loop) is similar for both cases. This emphasizes the importance of the CHUF protocol conceived by me and my collaborators.

We now consider what the same measurement protocols would show because of metastabilities associated with supercooling and superheating, if there was no kinetic arrest, as was discussed by us in detail in reference [8]. We first consider, using the schematic in figure 6(a), the isothermal M vs H curves. We reach (in zero field) a certain temperature $T_0$ (point Q) lying between $T^*(H=0)$ and $T^{**}(0)$. If we have reached $T_0$ by cooling (from point P to Q) in H=0, then the system is still in the FM phase that is in a metastable supercooled state. A subsequent isothermal cycling of the field at $T_0$ (Q to R to Q) takes the sample from a metastable FM state to a stable FM state and back to a metastable FM state, respectively. If however, we reach $T_0$ by heating (path S to Q), then the starting H=0 state is a stable AFM state. As H is raised, it transforms to stable FM at $H^{**}(T_0)$ but returns to metastable FM at H=0. The initial and final states at H=0 (at point Q) are thus different and we will observe an open hysteresis loop [8]. This open loop will not be observed when $T_0$ is reduced to below $T^*$ or raised above $T^{**}$. If the low T zero-field state is FM, as shown in the schematic in figure 6(b), then the anomalous open loop will be observed on reaching $T_0$ by cooling in zero field from P to Q (and not on heating, i.e. S to Q) provided $T_0$ is between $T^*(0)$ and $T^{**}(0)$. This experimental check for an open loop, on cooling versus on heating, thus identifies the low-T zero-H equilibrium state [8]. The open loop will not be observed as $T_0$ is lowered below $T^*$ in this case also. This latter feature contrasts with the case of kinetic arrest where the open loop becomes more prominent as $T_0$ is lowered. We need to measure at various $T_0$ to use such measurements to ascertain the presence, or absence, of kinetic arrest [8].

Let us now consider the use of the CHUF protocol when the system corresponding to figure 7(a) [figure 7(b)] is cooled in large H [small H] to a $T_1$ that is still above $T^*$ at that H. No transition will be observed, and we need to determine whether the system is supercooled or kinetically arrested. On lowering (raising) H to below (above) $H^*(T)$, the system will convert to the equilibrium state at T at the field marked by the arrow, and on warming there will only be one transition from the low-T equilibrium phase to the high-T equilibrium phase at $T^{**}$ (in both cases). In this case *no reentrant transition will be seen on warming in any constant H*. This is in contrast to what was explained using figures 4(b) and (c). Thus CHUF protocol has no significant feature in the absence of a glasslike arrested state. CHUF protocol provides a very convenient measurement that allows clear determination of the existence of a glasslike arrested state as against a supercooled state.

## Disorder broadening of the transition: manifestations of tunable phase coexistence

We now discuss the effect of small disorder on an underlying $1^{st}$ order transition. Early theoretical arguments of Imry and Wortis [9] showed that such samples would show a disorder-broadened transition, with the broad transition remaining $1^{st}$ order for small disorder. Since the correlation length for a $1^{st}$ order transition is finite, the transition proceeds through nucleation of finite regions of the second phase. It can be understood that these regions, of dimensions of the order of correlation length, can have slightly varying 'compositions' because of the small disorder. (Timonin [10] had considered a lattice divided into blocks having different transition temperatures.) These would have slightly varying transition temperatures (or transition value for the other control variable), resulting in a spatial distribution of the ($H_C$,$T_C$) line across the sample, and this spread of local ($H_C$,$T_C$) values across the samples would result in the ($H_C$,$T_C$) line being broadened into a band for samples with small frozen disorder. The first visual realization of such a local variation was provided by Soibel et al [11] for the vortex melting transition. A similar visual realization for an antiferromagnetic (AFM) to ferromagnetic (FM) transition, in Ru-doped $CeFe_2$, was provided by Roy et al [12] for the $1^{st}$ order transition being caused by variation of temperature (with field held constant), and also by the variation of field (with temperature held constant).

The major qualitative deviation in this phenomenology, from both the Ehrenfest classification and the modern classification, which is relevant for experiments is that this *broad $1^{st}$ order transition may not have an easily observable latent heat*. It is more specifically characterized by hysteresis, associated with supercooling and superheating, and this metastability-based hysteresis cannot be seen across a similarly broad $2^{nd}$ order transition. While it has been recognized historically that there are experimental difficulties in measuring a small latent heat and hysteresis in a physical property which varies sharply between the two phases can used to identify a $1^{st}$ order transition [13], this is a qualitative departure in experimentally establishing a $1^{st}$ order transition. This hysteresis should be seen with both control variables, and should be metastability-based (rather than due to the experimental artifact of a temporal lag), as can be checked by generating fluctuations in the metastable state [1].

We have argued above that regions, of dimensions of the order of correlation length, can have slightly varying 'compositions' and slightly varying transition temperatures. This results in the ($H_C$,$T_C$) line being broadened into a band. For the same reason, the spinodal lines corresponding to the limit of supercooling ($H^*$,$T^*$) and corresponding to the limit of superheating ($H^{**}$,$T^{**}$), would also be broadened into bands [14,15]. Each of these bands corresponds to a quasi-continuum of lines; each line corresponds to a region of the disordered sample with length-scale of the order of the correlation length. We have discussed earlier the scenario where (T*,H*) and ($T_k$,$H_k$) lines cross, bringing out the evolution from a metglass-like to an O-terphenyl-like situation. With (T*,H*) becoming a band, this evolution has an intermediate regime where $T_k$ at some H falls within the band. The 1$^{st}$ order transition is broadened and proceeds partially till $T_k$, when it is arrested or interrupted [see schematic figure 8]. We thus have the two phases coexisting, with the low-T phase growing until $T_k$. The two phases coexist with the fraction of the two phases being dictated by [$T_k$ – $T_L$]/[ $T_U$ – $T_L$] (which is obviously a function of the field H), and this coexistence persists all the way as T is lowered below $T_k$. This was the qualitative explanation we provided for the widely discussed phase coexistence observed across the 1$^{st}$ order FM-AFM transition in half-doped manganites. This was followed by Banerjee et al [15] to obtain continuously variable values of magnetization and resistivity, in various half-doped manganites, at the same value of H and (low value) of T. This behavior was shown by us in various classes of magnetic materials and has now been established across many 1$^{st}$ order magnetic transitions [16].

Before discussing how this affects the qualitative features observed, we must recognize for consistency that we are discussing regions with varying $T_C$; why should they have the same $T_k$? If such a kinetic arrest were to occur below a ($H_k$,$T_k$) line in the pure system, the disordered system would have a ($H_k$,$T_k$) band formed out of the quasi-continuum of ($H_k$,$T_k$) lines where each line would correspond to a local region of the sample. We thus have the schematics shown in figure 9, where each of the (H*,T*), (H**,T**) and ($H_k$,$T_k$) lines is now replaced by a band. The sign of the slopes of (H*,T*) and (H**,T**) bands is dictated by the requirement of Le Chatelier's Principle that isothermal increase of H must cause a transition from the AFM phase to the FM phase. The same argument (supported by some experiments [15, 17]), requires that an isothermal increase (decrease) of H should cause 'devitrification' of the arrested AFM (FM) phase to the equilibrium FM (AFM) phase.

With these schematics, we now discuss the qualitative features that would be expected under isothermal M vs H measurements, or under the CHUF (or CHUP, or CHUE) protocols, as disorder broadens T*, T**, and $T_k$ lines into bands. The modification for isothermal variation of H, in comparison to the discussion around figure 5, is only that the sharp transitions get broadened and occur over a range of H. This will not be discussed further. The modification to the CHUF measurements is qualitatively new; we discuss this in some detail.

First we look at the effect of the cooling field $H_{Cool}$. For $H_{Cool}$ > $H_2$ for the high-T phase being FM (and for $H_{Cool}$ < $H_1$ for the high-T phase being AFM), the behavior on cooling is again qualitatively similar to that discussed for the case of no disorder; except that the transition gets broadened and occurs over a range of T. Similarly for $H_{Cool} < H_1$ for the high-T phase being FM (and for $H_{Cool} > H_2$ for the high-T phase being AFM) the behavior on cooling is again qualitatively similar in that the 1st order transition is completely arrested and the high-T phase persists as a glass-like arrested state down to the lowest temperature. In the range of cooling field $H_{Cool}$ satisfying $H_1 < H_{Cool} < H_2$ we find that a fraction of the high-T phase transforms to the low-T phase while the remainder persists as a metastable arrested state. For the low-T phase being FM, the transformation persists over a range of temperatures that rises with increasing $H_{Cool}$, and the fraction that transforms also rises with increasing $H_{Cool}$. For the low-T phase being AFM, both decrease with increasing $H_{Cool}$. We now have a phase transformation (that will show hysteresis on warming in the same H) that is interrupted and ends in a state of phase coexistence, and this phase-coexistence persists without further evolution down to the lowest temperature. This was sometimes attributed to a new inhomogeneous ground state; we now understand that *a disorder-broadened 1st order transition is getting interrupted by glasslike kinetic arrest*. The cooling field $H_{Cool}$ can be used to tune the coexisting fractions, and since the system is now kinetically arrested, we can change H at low T without changing this phase fraction.

Just like a cooling field $H_{Cool}$ lying between $H_1$ and $H_2$ allows the transforming fraction to vary continuously, similarly a warming field $H_W$ lying between $H_1$ and $H_2$ allows the devitrifying fraction to vary continuously. The behavior is understood exactly similarly if the starting lowest-T state is the totally arrested state (i.e. the high-T phase does not transform at all during cooling, and H is changed to $H_W$ at the lowest T). The process becomes more interesting if both $H_{Cool}$ and $H_W$ lie between $H_1$ and $H_2$. We shall discuss below the case where the low-T phase is AFM; the discussion for the low-T phase being FM is analogous and will only be summarized.

We consider $H_{Cool} = H_3$ and $H_W = H_5$ with $H_3 > H_5$. We denote by $X_3$ and $X_5$ the fraction of the FM phase that would be arrested if we used $H_{Cool} = H_3$ and $H_{Cool} = H_5$ respectively, and note that $X_3 > X_5$. It follows quite naively that for the CHUF path with $H_{Cool} = H_3$ and $H_W = H_5$ one would observe devitrification of $[X_3 - X_5]$ fraction from arrested FM to equilibrium AFM as one warms across the $T_k$ band. The exact temperatures where this devitrification occurs depends on the width of $(H_k, T_k)$ band and on the details of the lines within; this has been discussed in our publications [14]. For another CHUF path with $H_{Cool} = H_4$ [$H_3 > H_4 > H_5$] and $H_W = H_5$ the fraction that devitrifies would be smaller. And for $H_{Cool} = H_6$ [$< H_5$] and $H_W = H_5$ one would not observe any devitrification. These are CHUF protocols where the warming field is kept fixed and various cooling fields are used. The measured property has to be drastically different for the two phases so that one can easily estimate the phase fractions; magnetization is the obvious choice. In figure 10(a) we show our data for $H_W = 4$ Tesla in a magnetic shape memory alloy NiCoMnSn sample showing glasslike kinetic arrest of the 1st order austenite-martensite transition that is accompanied by a sharp drop in magnetization [18]. We can also consider CHUF protocols where the cooling field is kept fixed, and $H_W$ is varied. Here the arrested FM fraction is fixed; the fraction that will devitrify rises as $H_W$ decreases. This is again depicted in figure 10(b) for $H_{Cool}$ = 6 Tesla, for the same sample.

Similar behavior will be observed for CHUF protocol for the low-T phase being FM, with devitrification occurring only for $H_{Cool} < H_W$. We show in figure 10(c) our observations for PCMAO sample for $H_W$ fixed at 4 Tesla, and in figure 10(d) our observations for the same sample for $H_{Cool}$ fixed at 2.5 Tesla [9].

## Potential of the new measurement protocol

One should note that the CHUF protocol gives a path-dependent evolution of the state, and is not restricted to a particular measurement. It allows the experimenter to study various propertied as the system evolves from a metastable arrested state to equilibrium, along paths controlled by the protocol. As has been noted by others investigating the tuning of coexisting phase fractions by cooling field in some doped cobaltites, "The CHUF experimental protocol, clearly showing the devitrification of the arrested state, thus gives unambiguous and rather visual evidence of the coexisting phases in the magnetic glass state" [19].

The CHUF protocol in which $H_C$ is fixed such that the high-T phase is fully arrested and $H_W$ is varied, allows the measurement of $T_k$ as a function of H in the region where $T_k < T^*$, a region that cannot be accessed in a controlled way while cooling. This measurement is not time consuming, and has allowed studies investigating the effect of slight changes in composition on the process of glasslike kinetic arrest. This has allowed postulates on the conditions that enable specific heat to be removed without removing latent heat, i.e. on conditions that favour glass-formation. In addition to providing detailed studies on phase coexistence, it could probably help address long-standing questions in glass physics. Finally, since the CHUF protocol allows controlled devitrification where temporal relaxation can be measured, this should prompt theoretical studies on how this non-equilibrium phenomenon proceeds. We shall discuss subsequently various physical processes that have been investigated, throwing new light, using the CHUF protocol.

We conclude by asserting that we expect very similar physics to emanate from studied on 1[st] order transitions in functional dielectric solids, with small inherent disorder, showing 1[st] order transitions under variation of electric field. We expect the phenomenology developed above to be of some use.

*The phenomenology described has been developed with many collaborators. I particularly acknowledge discussions with Sindhunil Roy, Alok Banerjee and Rajeev Rawat.*

*Since retired; Email: chaddah.praveen@gmail.com

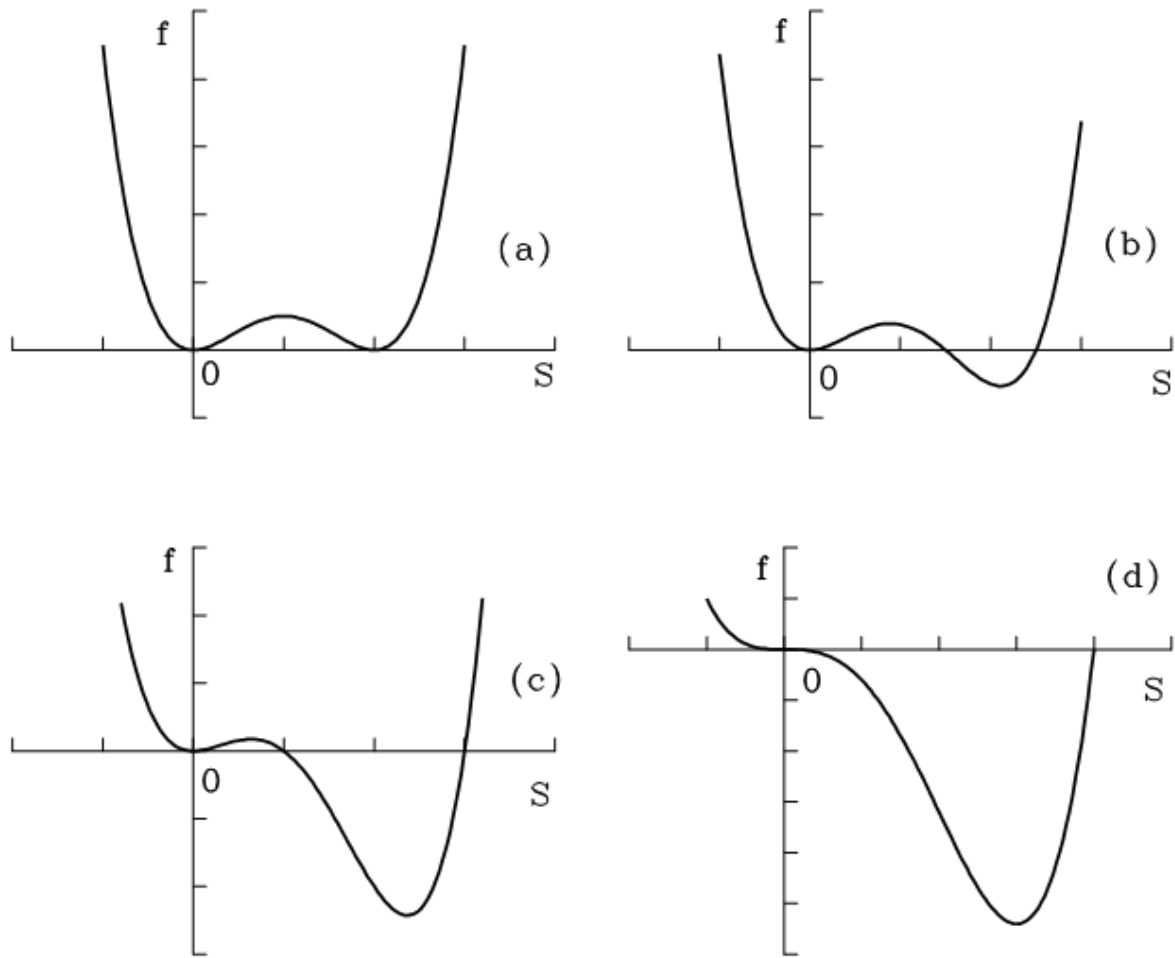

Fig 1: This schematic shows the free energy curves for various temperatures as a function of order parameter S, where the free energy has been expanded as $f[T,S] = a[T-T^*]S^2/2 - wS^3 + uS^4$. Supercooling is possible only till $T=T^*$, depicted in panel (d), where the barrier height reduces to zero. It would actually terminate when the height of the barrier surrounding S=0 reduces to about kT. See P Chaddah and S B Roy **Pramana-J.Phys 54,** 857 (2000); arXiv:9910437.

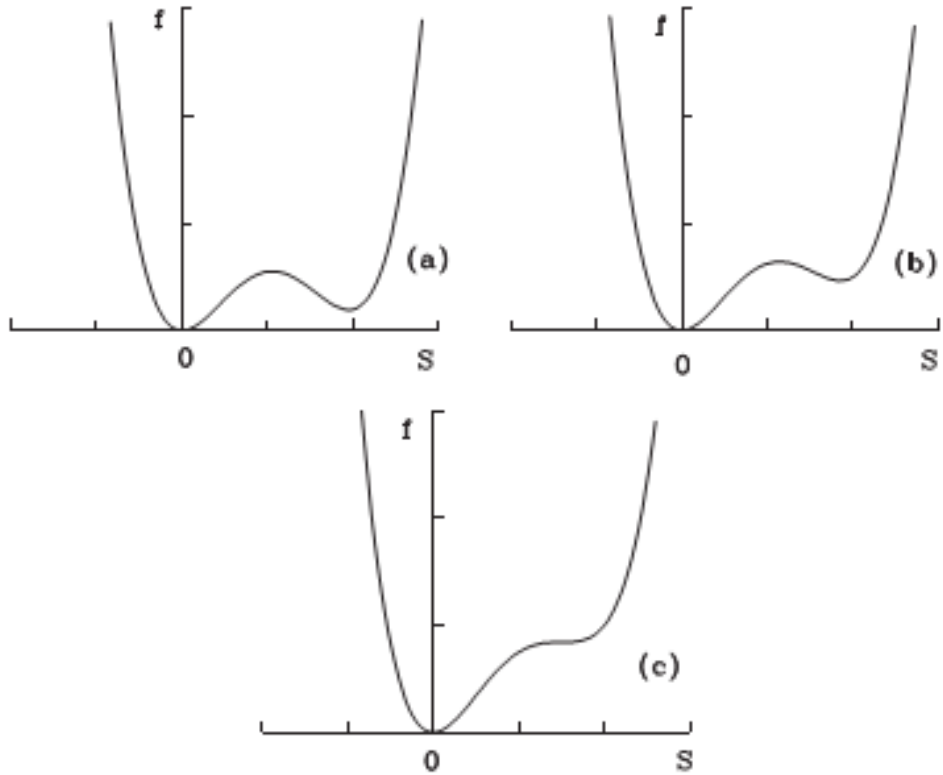

Fig 2: Schematic following that in figure 1. Superheating is possible only till T=T**= T* + (9w$^2$)/(16ua), depicted in panel (c). See P Chaddah and S B Roy **Pramana-J.Phys 54,** 857 (2000); arXiv:9910437. We wish to emphasize that f(S=0,T) is taken to be independent of T. If f is expanded in terms of a second control variable (say V), then f(S=0,V) cannot be assumed to be independent of V. This is naively obvious for the second control variable V being pressure P, since change in P will change inter-atomic distances and hence the energy.

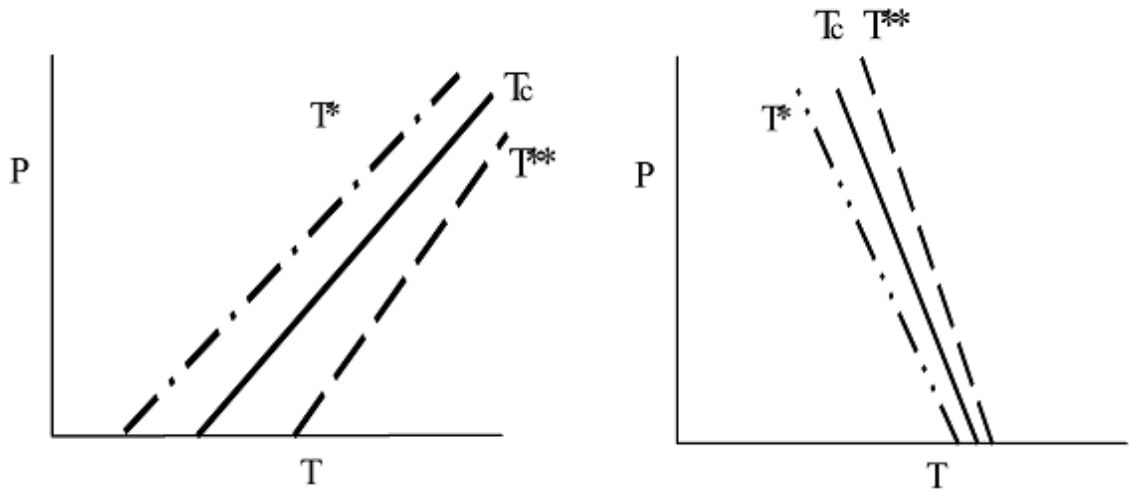

Figure 3: See P Chaddah and S B Roy **Phys. Rev. B60,** 11926 (1999). They discussed the phase transition under variation of only temperature at different pressures P, and the hysteresis given by the window between [T**(P) – T*(P)]. To quote from that paper, 'this window will increase with increasing pressure for the water-ice transition, and with increasing field in vortex-matter transitions. This window must decrease with increasing pressure for liquid-solid transitions in which the solid is more dense. These conclusions constitute a verifiable result.' Figure shown is from P Chaddah, **Pramana-J.Phys 67,** 113 (2006); arXiv:0602128. In this the conjecture was made for other second control variables. In this and all subsequent schematics the transition lines are assumed as straight lines. Introducing a curvature would not modify the phenomenology.

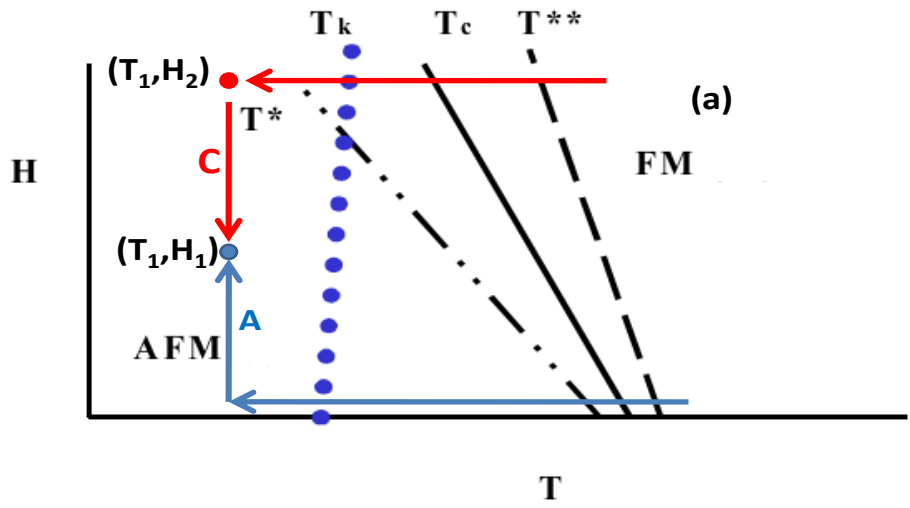

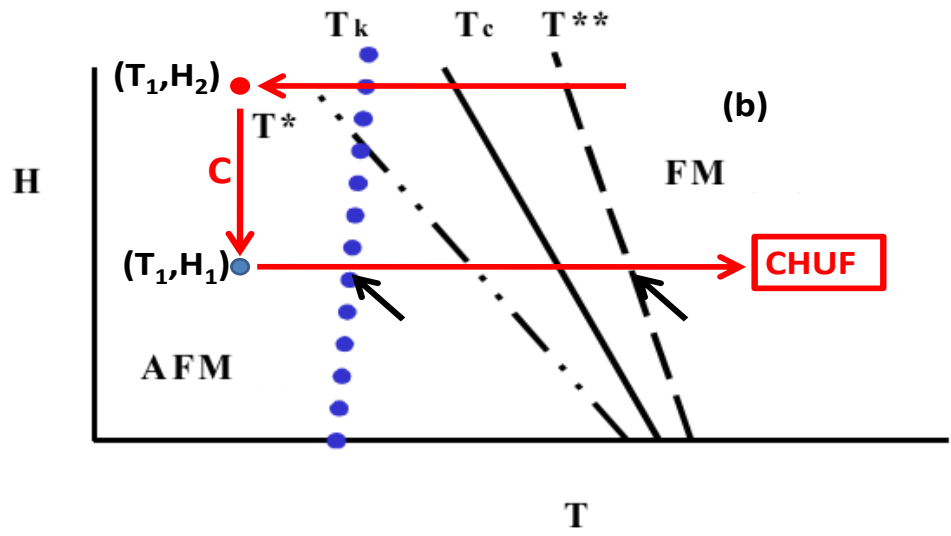

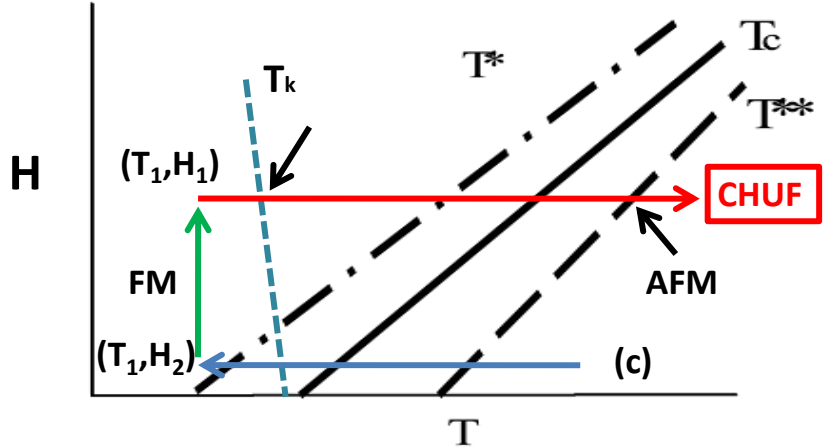

Figure 4: A 1st order transition from FM to AFM, with lowering T, is depicted in figures (a)&(b), while (c) is for AFM to FM with lowering T. Note that increasing H causes the AFM to FM transition. The path-dependence of the state at $(T_1,H_1)$ is depicted in (a) by following two paths. The CHUF protocol is brought out in (b) and (c) where the reentrant transition takes place on warming as the $T_k$ and $T^{**}$ lines are crossed, as marked by the black arrows. This reentrant transition is seen only when the cooling field is less than (more than) the warming field for the low-T equilibrium state being FM (AFM). This feature of 'devitrification' followed by 'melting' in the CHUF protocol would be seen even without disorder broadening of the 1st order transition. See P Chaddah, **Pramana-J.Phys 67,** 113 (2006); arXiv:0602128.

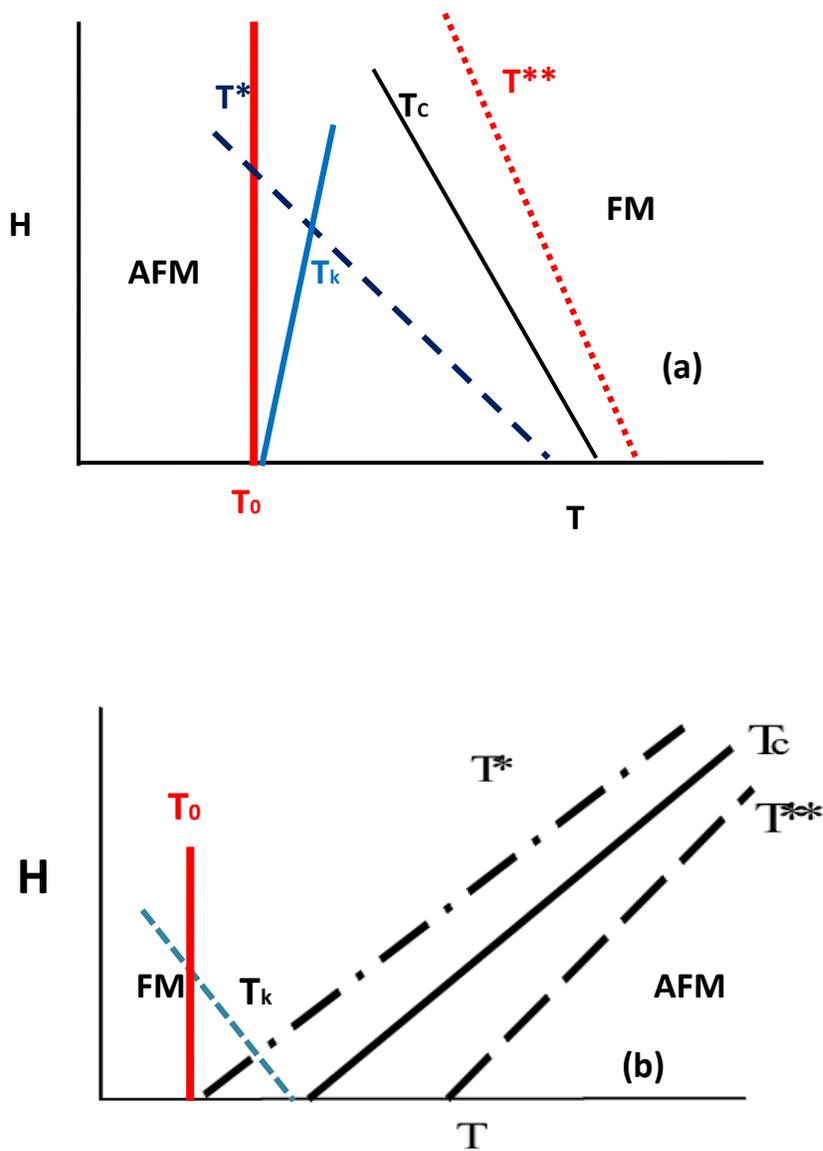

Figure 5: We consider in this schematic the isothermal variation (cycling) of H, after cooling in zero field, at a temperature $T_0$ chosen such that the $T_k(H)$ line will be crossed in this variation. Panels (a) and (b) correspond to the H=0 equilibrium state at $T_0$ being AFM and FM respectively. The initial state in both case is AFM, and the final remnant (H=0) state in both cases is FM. This measurement, by itself, does not allow a determination of the equilibrium state; see Banerjee et al [7].

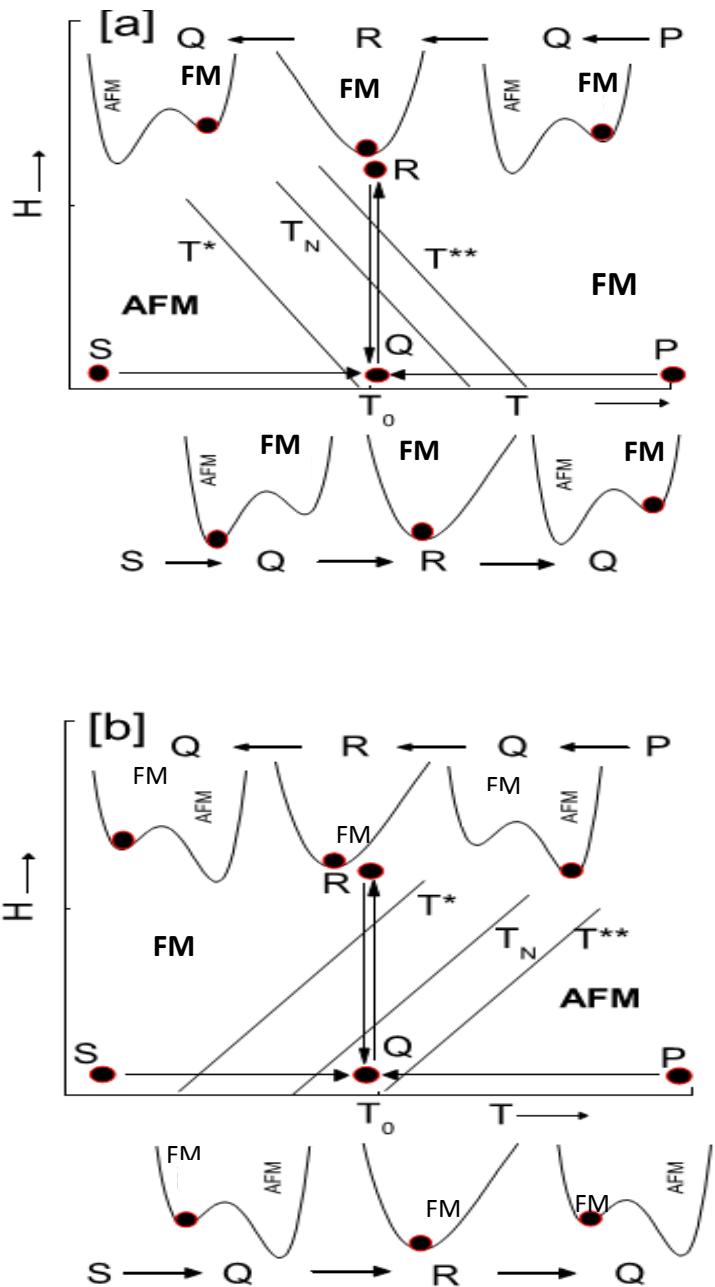

Figure 6: We show paths followed for an isothermal M vs H measurement at $T=T_0$ lying between $T^*$ and $T^{**}$. Schematics shown in (a) and (b) correspond to the low-T equilibrium phase being AFM and FM, respectively. An open hysteresis loop will be seen only if $T_0$ is reached on heating [on cooling] in case (a) [case (b)]. In these cases the initial state is an equilibrium state, while the remnant state is a metastable supercooled state. See P Kushwaha, R Rawat, and P Chaddah, **J Phys: Cond Matt 20**, 022204 (2008).

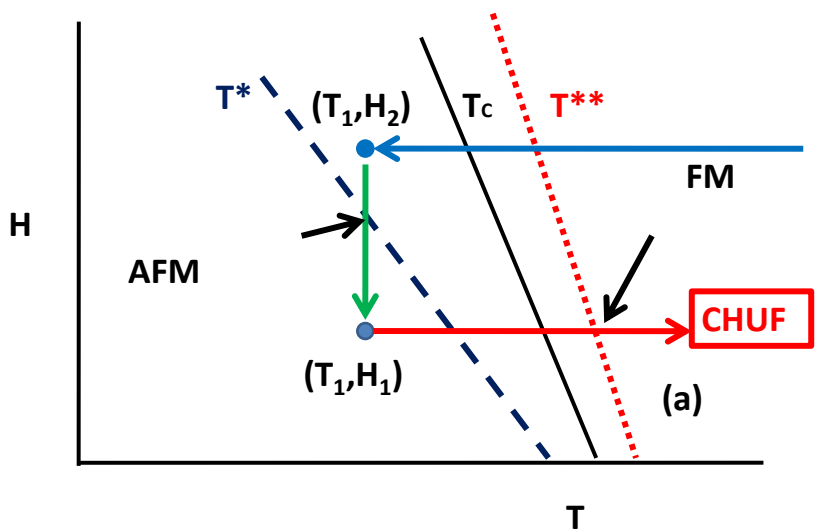

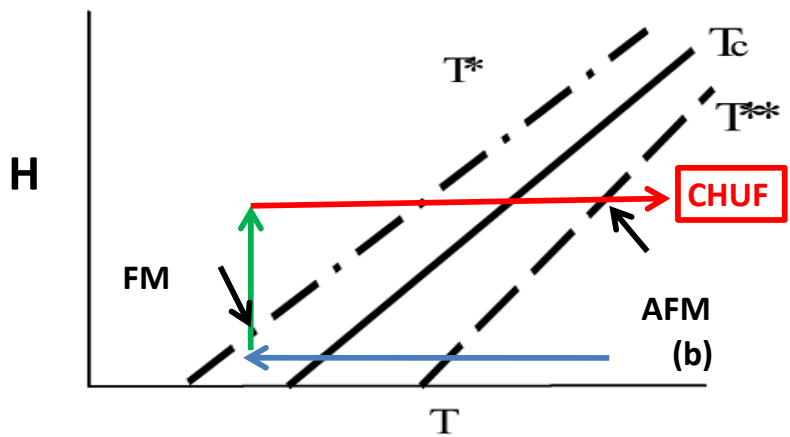

Figure 7: We consider a supercooled state, in which the phase transition is not observed at some H, but where here is no glasslike kinetic arrest. A CHUF measurement is done for the two cases of (a) an FM state being supercooled across the 1$^{st}$ order transition, and (b) an AFM state being supercooled across a 1$^{st}$ order transition. Following the CHUF protocol with the same sign of the inequality in cooling and warming fields, as in the corresponding schematics of figure 4, the warming process will not show a reentrant transition.

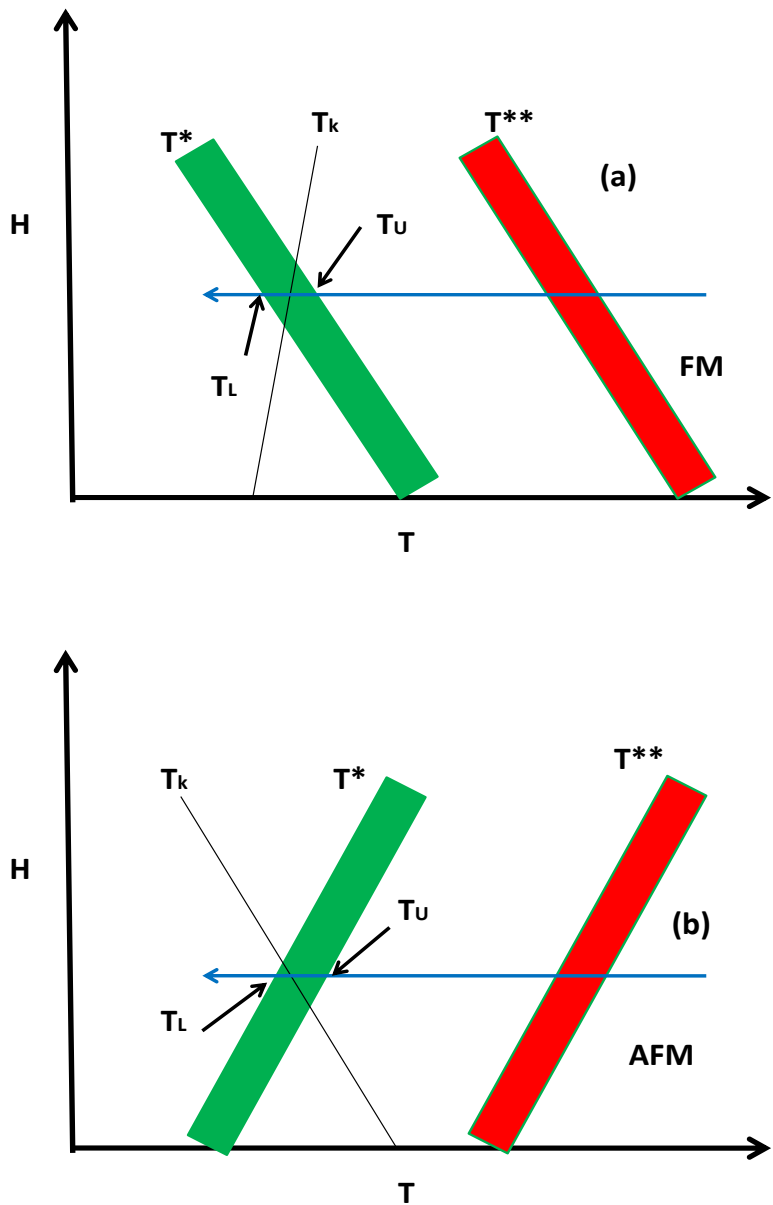

Figure 8: These are counterparts of the schematics in figures 4 and 5, but with each of T*, $T_C$, and T** being broadened into bands. We do not show the $T_C$ band in this and the subsequent figures for reasons of visual clarity; we are also assuming that the transformation occur at the limits of supercooling and superheating. Here the $T_k$ line overlaps with the T* band at some H, and the transformation is interrupted after it has proceeded in part i.e. up to $[T_U - T_k]/[T_U - T_L]$ (which is obviously a function of the field H). The arrested fraction, or the fraction of coexixting phases, can thus be tuned by the cooling field.

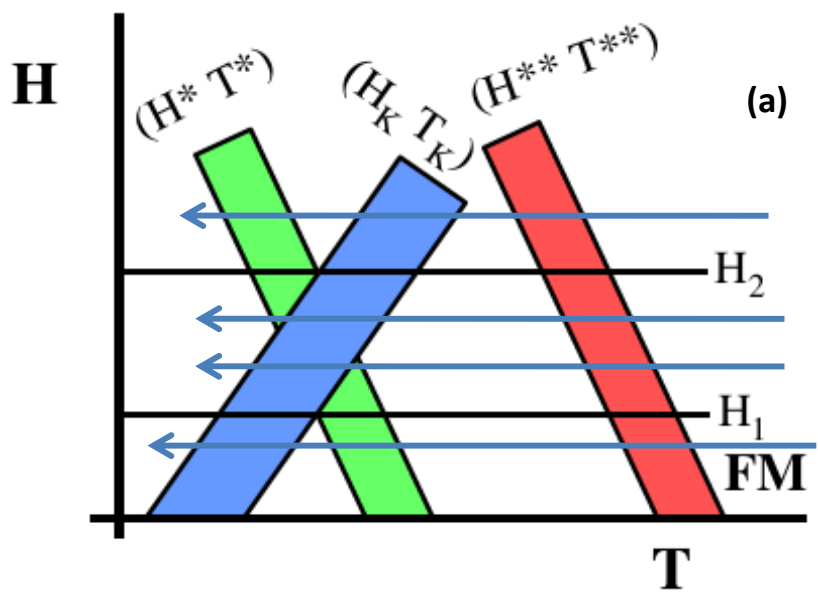

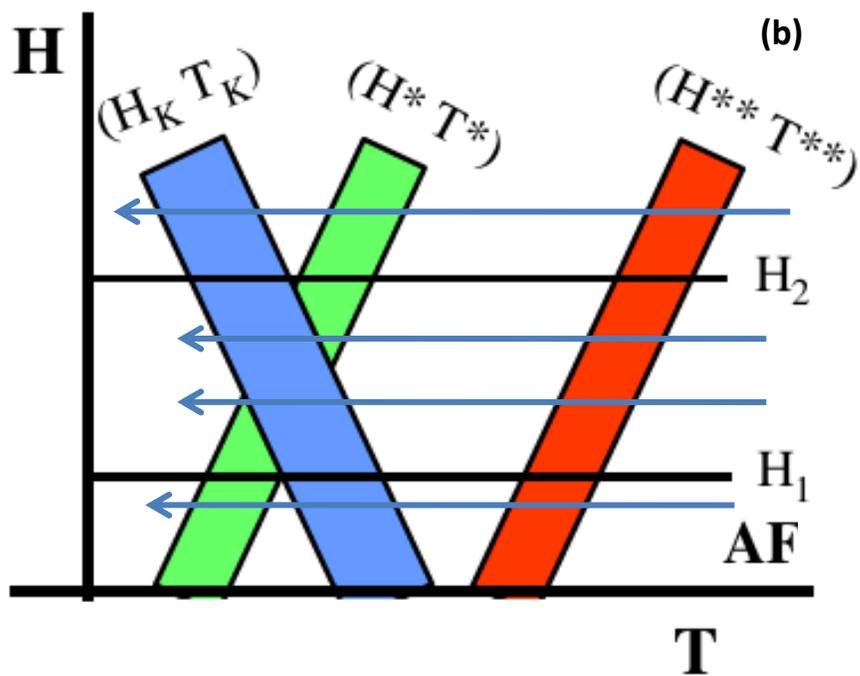

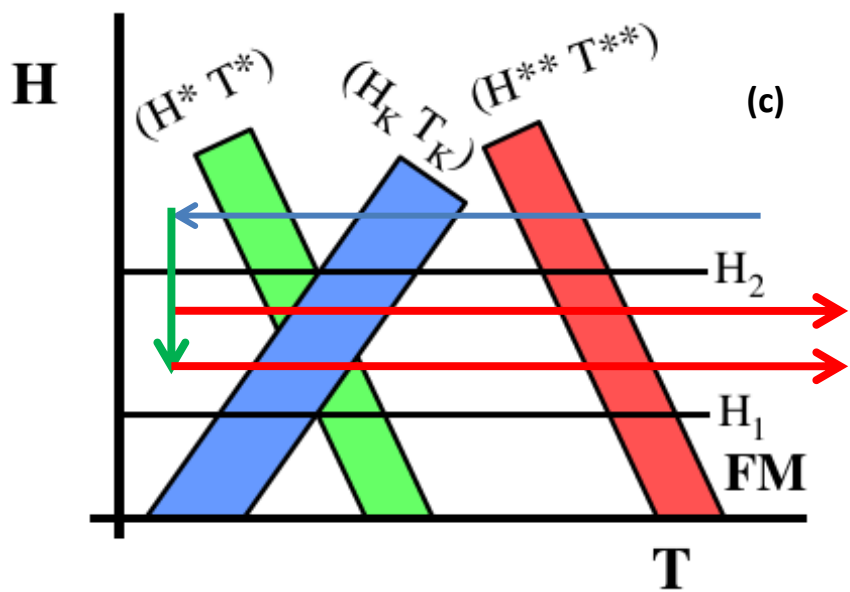
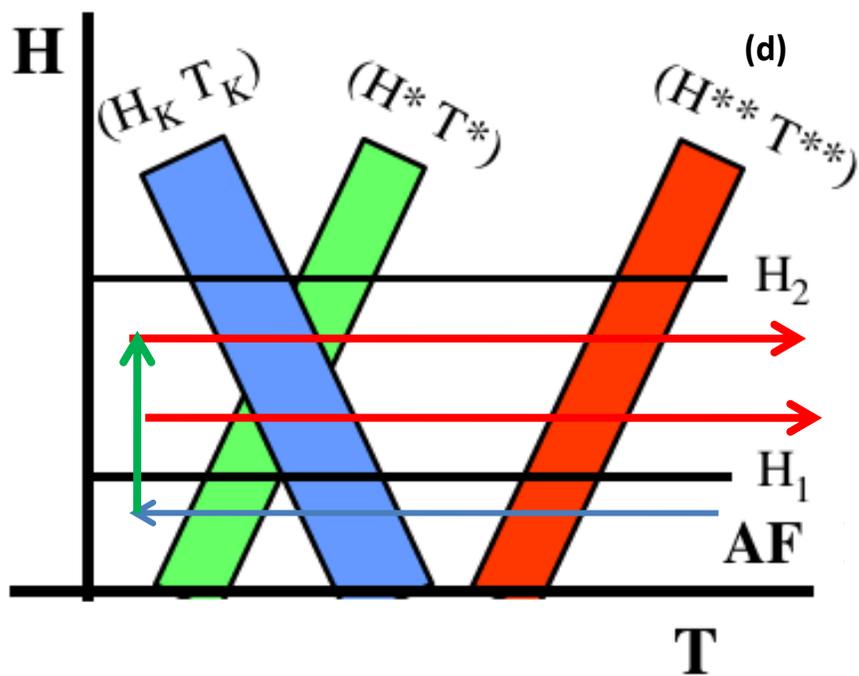

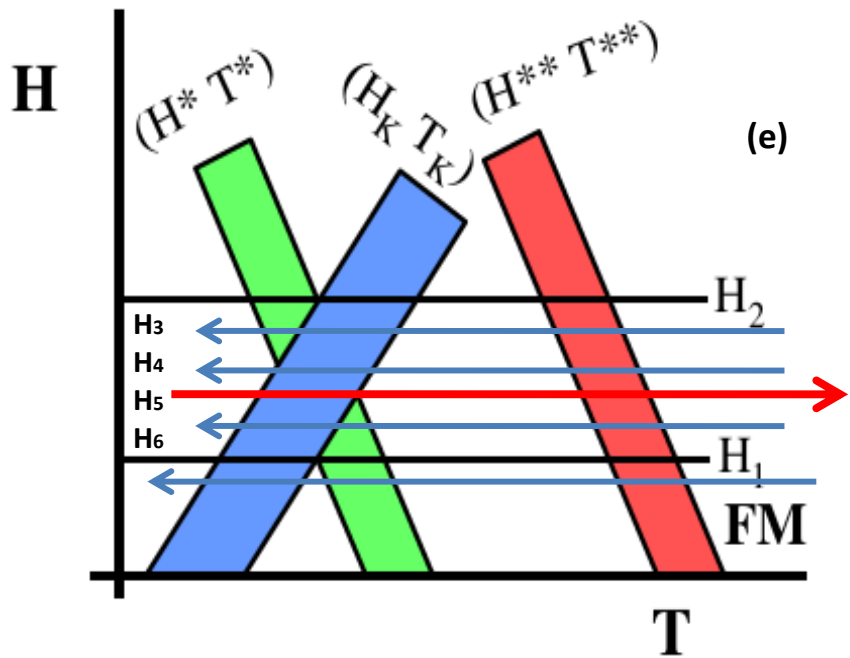

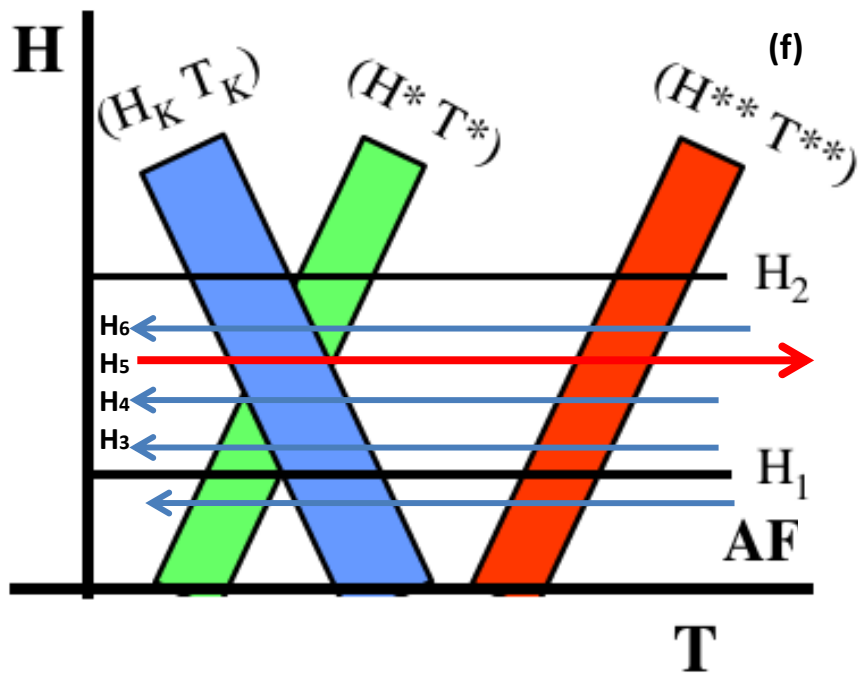

Figure 9: These are counterparts of the schematics in figures 4 and 5, but with each of T*, T** and $T_k$ being broadened into bands. Each of these bands corresponds to a quasi-continuum of lines; each line corresponds to a region of the disordered sample with length-scale of the order of the correlation length [14]. (Much work has been done to ascertain whether there is any correlation between the $T_k$ and T* of a region. As will be discussed separately, experimental

evidence suggests that regions with below average T* have above average $T_k$.) The blue arrows indicate different cooling fields, and the red arrows indicate different warming fields. In panels (a) and (c) the low-T equilibrium phase is AFM. It is totally arrested for $H_{Cool} > H_2$, and the arrested fraction increases continuously as $H_{Cool}$ is raised from $H_1$ to $H_2$, as depicted in (a). If we cool with $H_{Cool} > H_2$ to totally arrest the FM phase, and warm in different fields, then the devitrifying fraction rises continuously as the warming field reduces from $H_2$ to $H_1$ as depicted in (c). Panels (b) and (d) correspond to the low-T equilibrium phase being FM. Here (b) depicts that the arrested fraction decreases as the cooling field rises from $H_1$ to $H_2$, while (d) depicts that the devitrifying fraction rises continuously as the warming field increases from $H_1$ to $H_2$. Panels (e) and (f) depict the CHUF protocol with different cooling fields and a fixed warming field. Panels (c) and (d), on the other hand, depict the CHUF protocol with different warming fields and a fixed cooling field.

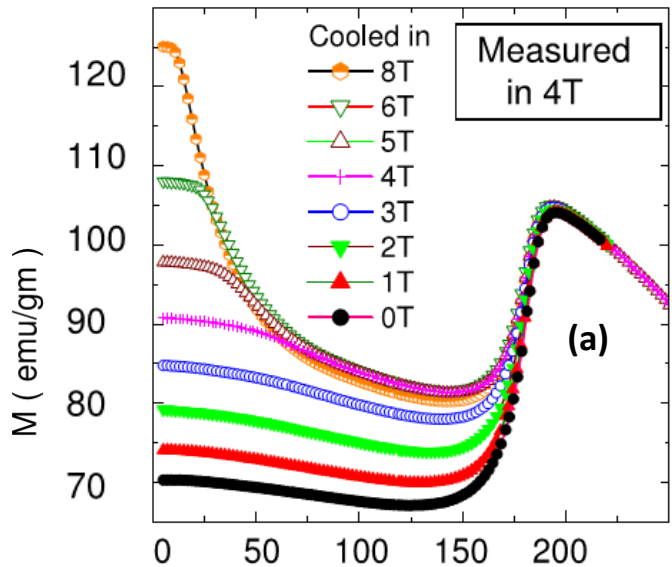

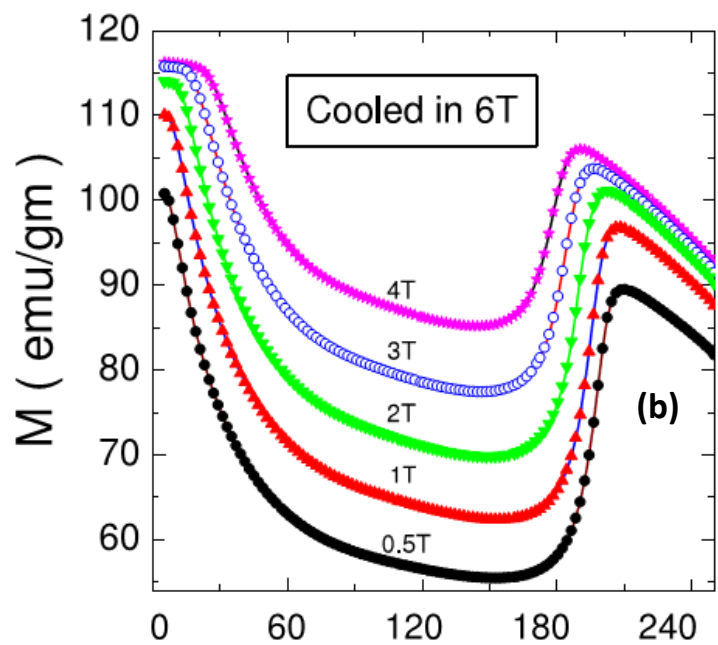

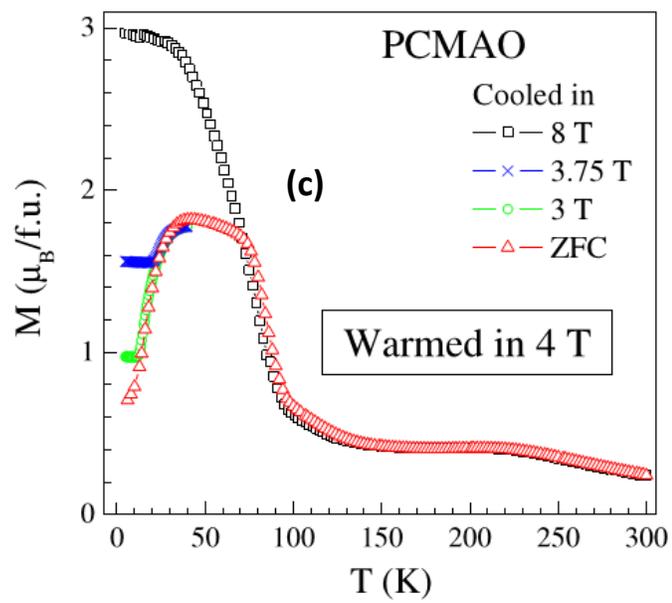

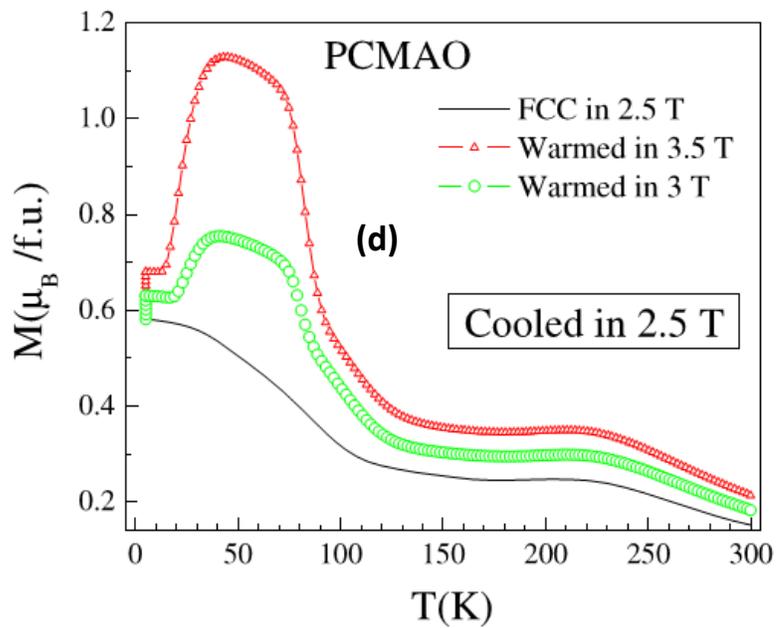

Figure 10: These are measurements of magnetization under the CHUF protocol in different samples. Panels (a) and (c) are for fixed warming field, while panels (b) and (d) are for fixed cooling field. Panels (a) and (b) are for a NiCoMnAl alloy, and data are taken from reference [18]. Panels (c) and (d) are for PrCaMnAlO sample, and data are taken from reference [9].